# Research Challenges and Prospective Business Impacts of Cloud Computing: A Survey

Amin Keshavarzi [1], Abolfazl T. Haghighat [2], Mahdi Bohlouli [3]

[1] Department of Computer Engineering, Marvdasht Branch, Azad University, Marvdasht, Iran. keshavarzi@miau.ac.ir
[2] Department of Electrical,Computer and IT Engineering,Qazvin Branch, Azad University, Qazvin,Iran. haghighat@qiau.ac.ir
[3] Department of Electrical Engineering and Computer Science, University of Siegen, Germany. mbohlouli@informatik.uni-siegen.de

*Abstract*— In today's information technology (IT) era, a major part of the costs is being spent on computational needs. Enterprises are in efforts to increase their Return on Investment (ROI) and individuals are trying to reduce their costs. In this regard, cloud computing which emerges as a fifth utility can reduce costs and enhance performance of IT solutions. A large number of companies and institutions are dealing with cloud related issues as a provider or user. Due to the fact that cloud computing services have been proposed in recent years, organizations and individuals face with various challenges and problems such as how to migrate applications and software platforms into cloud and how to ensure security of migrated applications and etc. Given that many different definitions of cloud computing is presented in many publications and projects, a concrete and clear definition for cloud computing considering its characteristics, models and services is provided in this paper. In addition, current challenges and open issues in cloud computing is discussed in details and further recommendations and roadmaps for scientific activities and researches as well as potentials for improvements in this area from scientific and commercial points of view are given in this paper.

## I. Introduction

Many changes have been occurred in the IT world from its very beginnings. Cloud computing is one the most important developments in this regard. The main idea of cloud computing has been proposed in 1960s. At that time, John McCarthy [1] said that computing will become a utility same to a telephone today. The practical implementation of the idea was not feasible due to the lack of appropriate infrastructure. In 1999, Salesforce launched its first cloud computing product. In 2002, Amazon corporation offered Amazon Web Service (AWS) and took an important step in order to implement the idea of cloud computing. Google, Microsoft and few others have then offered cloud based products and services to the market. Currently, many companies provide cloud based services and many others such as Japan Ministry of Economy [2] are using cloud based services to support their customers.

Public, private, hybrid and community clouds are four deployment models of cloud computing. In public clouds such as Blue cloud, AWS, GAE, Google Apps and salesforce.com, services are offered to general public. In private clouds such as Washington Trust and Indiana University, only authorized users of an organization can use services that are offered by the organization based on its resources. EMC, Eucalyptus and OpenStack are some of common tools for the configuration of private clouds. Hybrid clouds are the mixture of private and public clouds. In hybrid clouds such as Intel Hybrid Cloud, Real estate group and InterContinental Hotels Group, organizations can share parts of their owned resources with cloud providers and have also control on their own data and protect them against public cloud environment. Hybrid model has benefits of both public and private cloud deployment models. Different providers and/or users such as a group of hospitals in a city collaborate together in order to create a larger cloud environment entitled Community Cloud such as 3SO.

A brief history of cloud computing and review of selected definitions as well as suggested definition of cloud computing are given in the first section of this paper. Cloud service models are then studied in the section II. Essential characteristics of cloud have been proposed in section III. Key challenges and opportunities are described in the last section (section IV).

### A. History

As stated earlier, the idea of cloud computing was initially introduced by John McCarthy in 1961. He believed that one day the computation will become in the form of utility [1]. Douglas Parkhill posed the general computing problems in 1966 [3]. Due to the lack of proper infrastructure between 1960 and 1990, the idea of cloud was not feasible. Since 1990 with the advent of high speed networks, cluster was used to perform High Performance Computing (HPC). Cluster was a distributed system with a set of independent and connected computers used for large scale computations [4].

Foster et al has then proposed Grid technology in 1998 [5]. They defined Grid Computing as: "A computational grid is a hardware and software infrastructure that provides dependable, consistent, pervasive, and inexpensive access to high-end computational capabilities". The objective of Grid Computing [6] is the use of heterogeneous and distributed resources in an organization to build a powerful computer with high processing power [7]. In 1999, salesforce.com offered a service that would meet the goals of utility computing. Since then, many steps have been taken to achieve the goals of utility computing. Amazon Web Service (AWS), Google Application Engine (GAE) and Microsoft Windows Azure are some of basic steps in this regard.

*B. Current State and Statistics*

According to Gartner hype cycle [8], cloud computing was in the peak of inflated expectations stage in 2011. BitNami, Cloud.com and Zenoss released their survey [9] about key objectives for cloud adoption determined by over 500 IT professionals. According to their outlook,

- 7% of the people had an approved cloud computing strategy,
- 20% with no plan to develop a cloud computing strategy,
- 20% with a cloud strategy and cloud computing implementation,
- 22% with a partial strategy and
- 32% with efforts in collecting inputs for cloud computing strategy.

Development and test labs were the target goal for 61% of the overwhelming use-cases specified by surveyed participants. Accordingly, 37% aimed to offer software as a service, 33% to mimic public cloud computing capabilities behind a firewall, 27% for high performance computing (HPC), and 16% to resell hosting services [9]. The study of 451 Group with 417 responses garnered from both vendors and end users [10] shows that 40% of people are experimented with a move to cloud in 2011 and 26% wait for market maturity before cloud adoption. The study of AFCOM [11] about the poll of 358 IT managers and other senior IT executives indicates the growth of cloud computing as a trend for shaping data centers and shows that 36% of datacenters used cloud computing for their implementations in March 2011. This figure was 14% in October 2009 which shows the growth of clouds in data centers as well.

Computer Discount Warehouse (CDW) identified cloud computing [12] as a technology that helps organizations to make progress toward their key goals such as "consolidate IT infrastructure", "reduced IT energy" and etc. In response to the question, "What formal goals does your IT department have for the next two years?", made by CDW between 1200 responders, 42% selected "consolidate IT infrastructure" as their key goal, 42% selected "reduced IT energy". The term "enable or improve anywhere access" was the key goal of 38% and "reduced IT capital requirements" was the aim for 37%. IT directors believe that cloud based IT operation cost (people and tools) slightly less than not-cloud hosted applications [13]. In earlier stated study of the 451 Group, 55% of respondents believe that cloud computing has lower total costs of ownership. The total market size for cloud computing in 2012 is $40.7 billion. This figure will be $150 billion by 2013 according to Gartner which shows two years Compounded Annual Growth rate (CAGR) of 9.45 percent in clouds market.

*C. Current State and Statistics*

According to Forrester research [14] cloud market size will reach $241 billion by 2020. Figure 1 shows predicted global public cloud market size from 2011 till 2020. Kelly Livesey [15] predicts total revenue of $66 billion in 2016 for global public cloud services market. Software as a service (SaaS) will shrink from 87% of the market in 2011 to 62% in 2016, similarly infrastructure as a service (IaaS) from 9% to 23% and platform as a service (PaaS) from 5% to 16% by 2016.

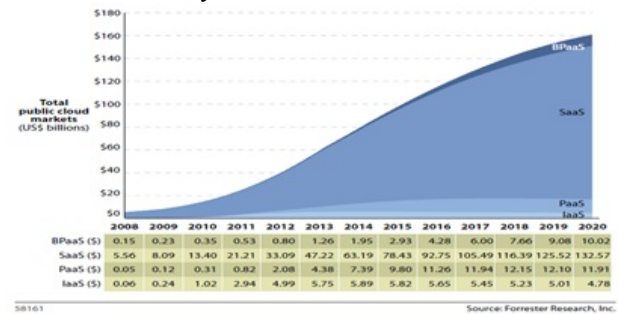

Figure 1. Public cloud market size from 2011 until 2020 [14]

While 65% of 225 people including IT and business executives, industry analysts and news media who registered for the cloud computing conference in Silicon Valley and responded to the survey [16] about future of cloud computing believe that cloud computing will be all out reality by 2015 and 62% of responders think that most CIOs will deploy only non critical applications in public cloud during 2011 till 2014.

## II. RESEARCH CHALLENGES AND ROADMAPS

The study of this paper about research challenges and future roadmaps of cloud computing are categorized into six groups including security, Autonomic Resource Management (ARM), adoption, development and benchmarking, Big Data, and Social Cloud. Providing secure cloud platform depends on the security of data, transmissions, applications and third party resources [17]. Target groups of threats, hackers (people who can make problems) and anti-hack strategies investigate security challenges. Traditional resource management techniques do not work efficiently for heterogeneous and large scale infrastructure of cloud resources and services in order to guarantee important Service Level Agreement (SLA) parameters, energy efficiency, scalability, fault tolerance and increased efficiency. Using cloud deployed services has some benefits from one side and some disadvantages and risks from another side. In making decisions to migrate or not, subjects such as cost and risk management are taken into account. An efficient and informative cloud benchmarking tools which can test a set of various workloads for instance on the cloud infrastructures (IaaS) to summarize and visualize detailed technical measures and reports with values such as response time, service up and down time, costs, performance, utilization and etc in order to compare different providers are still missing. The big data requires 2 type of data management system for update and analyze of data. Cloud computing characteristics make it attractive for both types of data management systems: update heavy applications and analytics and decision support systems [18]. Cloud computing and social media have characteristics that are complementary. These technologies can integrate as two ways: using cloud computing in social media deployment and using social network in cloud computing deployment.

### A. Security Challenges of Cloud Computing

Pearson et al [20] define privacy as "For organizations, privacy entails the application of laws, policies, standards and processes by which Personally Identifiable Information (PII) of individuals is managed". Due to shared environment of cloud computing, remote access and process of the data and combined service and information flow across provider boundaries, privacy is one the main considerations in cloud computing [46]. Identity and access management is one of the mechanisms that are used to preserve privacy. In access management, access to various resources and services are controlled through mechanisms such as authentication and authorization mechanisms. In authentication, identity of applicant is verified and his/her access level is controlled in authorization step. Cloud environment consists of multiple services and domains and each domain has own access policy. It is needed to design an access control framework that integrates access policies of multiple domains. Security Assertion Markup Language (SAML), Extensible Access Control Markup Language (XACML) and web service standards are various specification frameworks for cross-domain access specification and verification [21]. Identity federation is the way that organizations and public cloud providers trust each other and share digital identity and attributes and with this manner, support single sign-on [22]. SAML and OpenID standards can be used to accomplish identity federation [22]. XACML uses a XML-base language to define policies and decision making task. Identity Management (idM) method has been used for access control in [17]. Role Base Access Control model (RBAC) is proposed in [23] to control APIs as interfaces between customers and provider.

Traditional integrity checking techniques such as hashing cannot be applied to data and computation integrity in cloud computing, because hashing of such huge volume of data through internet is not feasible. Provable Data Possession (PDP) can support integrity check in cloud computing. In order to preserve confidentiality, traditional techniques cannot be applied to cloud computing because of existing threats inside the systems. In confidentiality method privacy of users is protected from others and authorized users can only access to the environment. Virtualization helps to separate shared environment of users from others and isolate them. Subashini and Kavith [17] used security protocols such as SSL or IPSec in encryption phase to prevent security threats in clouds. There are two types of threats [24] for cloud service availability: flooding attack via bandwidth starvation and Fraudulent Resource Consumption (FRC) attack. FRC attack detection and defending the new DOS attack [24] can repel these attacks. In some other models a combination of earlier stated techniques is used. For example, Zissis and Lekkas [25] used Trusted Third Party (TTP) as an entity which both parties trust on a third protocol to provide a secure interaction. In this model, a set of TTPs create a Public Key Infrastructure (PKI) and are responsible for ensuring the security of the cloud environment. PKIs with a directory of certificates are used in order to provide an access control. PKI is also used with Single-Sign-On (SSO) which the user does not need to sign in repetitive actions.

As a summary, authors believe that security is the top challenge of cloud computing and there is a lack of new cloud specific techniques and methodologies in connection to security and trust on cloud computing. Researchers and service providers need to offer novel and smart techniques and algorithms to address following questions.

- How do cloud providers secure the message exchanges as well as agreements and transits of data and applications during the migration to cloud and inter-clouds?
- New and cloud-specific algorithms to verify user identities in order to protect a system from security threats?
- How cloud providers can predict attacks to the system before happenings and incidents and how users can contribute to cloud services in order to rapid detection of data leakages and protect data leakage?
- How to develop an indicator system to monitor and measure security performance of cloud offered services from different vendors to choose the most fitting secure solution based on the target goals of clients?
- Which techniques can be used in order to preserve privacy in multi tenant cloud computing environments?

### B. Autonomic Resource Management(ARM)

There is still a lack of autonomic and intelligent methods in managing heterogeneous and large scale infrastructure of cloud resources and services to guarantee Service Level Agreements (SLAs) and provide energy efficient, scalable and fault tolerant services. SLA is a contract between customer and provider including some criteria about Quality of Services (QoS). Provider has to observe SLA guaranteed criteria and has to pay penalty to the customer in case of any failure or descent from agreed SLA. Therefore, the major issue in ARM systems is about automatic prediction of failures and violations before happening in order to automatic balance of the current workload, incoming requests and available resources. Knowledge Management (KM), Artificial Intelligence (AI), machine learning and Decision Making (DM) techniques can benefit dynamic monitoring and intelligent management of resources for increased efficiency. Maurer et al [26] used KM techniques in ARM system and allocated physical resources to VMs for detecting SLA violation. Case Base Reasoning (CBR) has been used for DM in Monitor, Analyze, Plan and Execute (MAPE) cycle of an autonomic SLA management. The monitoring phase consists of two components: host and run time monitor. Low level metrics of the infrastructure's resources are being first monitored and then mapped to high level SLA parameters for SLA violation prediction. Analyze phase is responsible for detecting SLA violation. Planning phase maps knowledge based suggested actions to PMs, prevents oscillations and schedules execution of tasks. Simultaneously, Emeakaroha et al [27] mapped low level metrics such as downtime, uptime and available storage to the agreed SLA quantities using mapping rules that exist in KB. In addition, KB helps ARM system to detect SLA violations before happening. Their system consists of an application deployment component which manages execution of user applications and an automatic VM deployment to allocate required resources to requested services and arrange its deployment in VMs.

Gulati et al [28] have investigated some challenges in building resource management system in clouds. According to this research, when the size of a cluster increases, management space decomposition is needed in order to decrease resource provisioning time. Adding new machines over time and heterogeneity of clusters in cloud computing is another explored challenge in this research. Increasing frequency of operations and impacts of failures on resource management components are also discussed as resource management challenges in clouds. Using intelligent techniques for ARM systems in clouds which consists a set of large-scale data centers with huge energy consumptions and expensive maintenance costs can also decrease energy and maintenance costs and air pollution with respects to green computing solutions. Buyya et al [29] has proposed following methods for decreasing energy consumption in clouds in their visionary paper:
- Design new architectures for data center with respect to energy consumption.
- Develop efficient resource allocation mechanisms and scheduling algorithms to decrease energy consumption.
- Design and development of software components for energy management in clouds.

Dual Agreement Protocol of Cloud-Computing (DAPCC) [30] can decrease the number of message exchanges to reach to an agreement on public criteria in SLA, which tolerates the maximum allowed number of error components. Low Latency Fault Tolerance (LLFT) middleware [31] that use leader/follower replication approach makes distributed applications fault tolerant in cloud environment. LEFT consists of Low Latency Messaging Protocol (LLMP), a Leader-Determined Membership Protocol (LDMP), and a Virtual Determinizer Framework (VDF). LLMP is a reliable message delivery service. LDMP is a fast reconfiguration and recovery service for faulty replicas as well as joins or leaves to the group. VDF ensures that the order of information is same in main version and backedup version. Higher replica consistency, application transparency and low end-to-end latency can be achieved in LIFT [31].

As discussed, autonomic resource management in cloud computing with respect to the energy efficiency, green computing, service level agreements, quality of services and fault tolerance is one of the main challenging issues in cloud computing. Researchers and service developers need to consider following questions in order to provide an efficient ARM system in Clouds:
- How to adopt traditional knowledge management techniques for predictive failure detection in clouds?
- Propose innovative ideas of using cloud computing for non-IT areas such as heating of building?
- Novel algorithms to efficient, green and optimized power consumption in clouds?
- Autonomic self-* cloud resource management services?
- Adaptive indicator system in ARMs for autonomic evaluation of resource management services?
- Potentials to increase fault tolerance of clouds by ARMs?

*C. Cloud Adoption*

Decision-makers and Chief Information Officers (CIOs) should evaluate their benefits and risks to migrate and adopt cloud by means of decision making tools. In addition to the essential decision to migrate, they also need to assign proper deployment model, service model and provider. There is a lack of an intelligent system to assist CIOs to measures based on the type of aimed workloads, required security level, response time and etc to find proper service and deployment model. Paul Bannerman [32] categorized cloud adoption risks from different points of view including scientific, practical and industrial views into 10 categories as follows: Security,

Lock-in, Control, Legal, Service, Performance, Cost, Governance, Competencies and Industry.

TABLE I. RISK FREQUENCIES FOR ADOPTION OF CLOUD COMPUTING FOR ENTERPRISES [39]

| No. | Risk category | Practice % | Practice Freq. | Research % | Research Freq. |
|---|---|---|---|---|---|
| 1 | Security | 23.8% | 104 | 38.4% | 38 |
| 2 | Lock-in | 22.9% | 100 | 22.2% | 22 |
| 3 | Control | 13.3% | 58 | 8.1% | 8 |
| 4 | Legal | 12.8% | 56 | 11.1% | 11 |
| 5 | Service | 11.0% | 48 | 9.1% | 9 |
| 6 | Performance | 5.0% | 22 | 5.1% | 5 |
| 7 | Cost | 5.0% | 22 | 4.0% | 4 |
| 8 | Governance | 3.7% | 16 | 2.0% | 2 |
| 9 | Competencies | 1.4% | 6 | 0.0% | 0 |
| 10 | Industry | 1.1% | 5 | 0.0% | 0 |
| | | 100.0% | 437 | 100.0% | 99 |

According to the Table 1, security is the highest risk for enterprises. Technology Suitability Analysis, Energy Consumption Analysis, Stakeholder Impact Analysis, Responsibility and Cost Modeling are 5 useful components developed in cloud adoption toolkit in [33] to support adoption decisions of enterprise.

As a result, researchers, cloud service providers and developers should consider further studies and improvements in the following issues and challenging questions:

- Economical and technical measures about when, how, where and which cloud services are needed.
- Identify the best cloud provider and service model based on the specific application domain of an enterprise?
- Customize the cloud hosted models and integrate with the traditional in-house systems and solutions?

### D. Cloud Development and Benchmarking

Different users from wide variety of domains and various workloads use cloud services. Benchmarking tools should support multi-domains and could customize measurement variables based on domains. Cloud providers and researchers need to first develop a benchmarking method and metrics including all relevant indicators as well as proposing the helpful tools in order to adopt with these indicators. Yahoo! Cloud Serving Benchmark (YCSB) [34] is for evaluation and comparing PNUTS system with other cloud Database (DB) systems. Performance, elasticity, availability and replication are some of indicators used in this tool. This tool is limited to the Cloud DB systems. Schad et al [35] used CPU, I/O and network variants as indicators to evaluate the efficiency of Amazon EC2 and compare it with local cluster systems. Kossmann et al [36] used transaction processing to investigate various business service that use different architecture for database applications. Specified metrics and workload in a CloudGauge as a dynamic and experimental cloud benchmarking suite [37] with the main focus on performance evaluation of virtual systems can be used for performance models of virtual systems and clouds. Due to the wide variety of cloud services and providers, developing of an efficient integration suite for cloud hosted tools and package is a basic task in cloud development models. Integration should be done from various points of view such as computation services and storage services provided by various providers [38]. In this regard, an integration of in-house and cloud hosted resources should be supposed for integration framework.

As a result, existing programming models are not sufficient in order to support cloud potentials and developing new programming models and paradigms is needed. Cloud service brokerage (CSB) is an intermediate between providers and consumers [39] which help users in finding and managing service consumptions. Users can install on their desktop computers and use broker as a service [40]. CSB primary roles have been noted as aggregation, integration and customization in [39]. Nair et al [41] proposed a secure cloud broker architecture that can be used to implement brokering of multiple providers and provide a SLA-based pricing model to broker users. This broker has the following capabilities: data confidentiality, scaling resources, identity and access management, risks analyzing, cloud bursting, securely consumers' data transferring, SLA management and so on. Scientists and service providers should address following points by proposing new techniques and methodologies:

- Transparent and reliable message passing and interoperability interface between VMs and physical machines in the same or separate clusters?
- Mass-customized and adaptive programming languages for inter-operability VMs between different vendors.
- A multi domain workload generator to evaluate various cloud services and models and visualize the results.
- Semantic and intelligent performance evaluation tools

### E. Big Data Management and Cloud Computing

The world of information is doubling every two years [42]. According to Cisco VNI, IP traffic will reach 966 Exabyte by 2015 [43]. IDC says that digital universe will be 35 zettabytes in 2020 [44]. There are three different dimensions for big data namely "Volume", "Velocity" and "Variety" [53]. Big data is defined in Wikipedia as: "Big data is a collection of data sets so large and complex that it becomes awkward to work with using on-hand database management tools. Difficulties include capture, storage, search, sharing, analysis, and visualization".

Chaudhuri et al [48] defined 6 challenges about big data and cloud computing as data privacy, data accuracy, and data exploration to enable deep analytics, enterprise data enrichment with web and social media and query optimization. Traditional solutions for management of big data are not appropriate to be used in clouds due to cloud characteristics. Design a novel cloud deployed big data management systems that fit for cloud facilities is a hot research topic. Having a decision support system (DSS) [47] and machine learning for large scale disparate data sources that can be delivered as a service to consumers is another main research challenges in this area. For this purpose, DSS should be developed in a component base approach that can be characterized by reusability, substitutability, extensibility, scalability, customizability, reliability, low cost of ownership, economy of scale [45]. Web applications have different storage requirements than traditional applications and data stores (Relational Database Management Systems, RDBMS) do not reply to those. Strong consistency and integrity are not vital for web application instead low latency, distribution and scalability is important. According to CAP theorem, a distributed system can satisfy only two of the following three requirements: consistency, availability and partition tolerance [49]. The NoSQL (interpreted as Not Only SQL or Not relational) data stores have some characteristics such as schema free, easy replication support, simple API, eventually consistent/BASE (not ACID), a huge amount of data and so on [49]. In [49] a list of NoSQL databases has been depicted. Cattell et al [50] have compared some NoSQL data store. Cloud storage has some challenges such as security, control, performance, support, configurability and vendor lock-in and so on. In [51] a framework has been offered to enable the execution of large scale data mining applications on top of cloud computing services. This framework has been developed using windows Azure and consists of 5 components: a set of binary and text data containers, a task queue, a task status table, a pool of workers and a websites [51].

Demirkan et al [45] have proposed a conceptual model for evaluation of service-oriented decision support systems (SODSS). In this paper architecture for SODSS has been offered that consist of 3 layers: IaaS, SaaS and business process (BP). In a SODSS there are 4 major components: information technology, process, people and organization. Also operational systems, data warehouses, online analytic processing and end user components can be delivered as a service to users. In Data as a Service (DaaS), data can reside anywhere and business process can access 24 hours a day. Information as a service (IaaS) provides a fast access to information across a business for users and processes. IaaS offers an integrated platform of information that provides a set of interfaces and standards to access data easily. In Analytics as a Service (AaaS), cloud computing is used for analytic work which is also called Agile Analytics. Scalability and cost reduction are benefits of using Agile Analytics. Encrypted data analyzing and using data mining algorithms for big data are challenges of Cloud Analytics [48]. As a summary, authors suggest to scientists and service providers to consider developing new mechanisms and algorithms considering following points:

- Scalable decision making algorithms for big data analytic in cloud computing?
- Scalable data management and machine leaning techniques in clouds?
- Adapt traditional machine learning, data mining, decision making and artificial intelligence methods and algorithms with cloud hosted scalable algorithms and methodologies?

F. *Social Clouds*

Social media needs the scalable infrastructure such as clouds due to its huge number of users. In order to prevent confidentiality violation, providers can share physical infrastructure between a customer and her friends that exist in friend list. Chard et al [54] have proposed the social cloud architecture which uses Facebook as a foundation for user identification. A banking component has been used for reservation and credit transfer between users. This architecture has a component named service marketplaces that supports two types of markets:

- In posted price model, when a user requests a service, the social cloud will generate a list of services with their attributes and options such as availability and price by means of offerings for friends of the target customer in social media. User will then select his/her desired service and SLA which is created in social cloud. SLA will then send to banking component in order to transfer credit between users.
- In auctions model, users will first define their requirements and will submit to social cloud infrastructure. Each provider will create a bid based on submitted requirements. Finally, auctioneer will determine the winner and will create the SLA between auction initiator and the winner. Same to the first model, the SLA will be sent to banking component in order to manage agreed credits and service costs.

Li et al [52] have demonstrated a social service computing ecosystem. Their ecosystem has five basic elements including: service providers and consumers, services, local services, physical things and cloud computing platforms. In this ecosystem, there are four types of networks such as social networks (between service providers and service consumers), service networks, cloud computing networks and physical thing networks.

As a summary of this challenge authors believe that service networks, basic social service computing infrastructures, social service publishing, social service discovery and ranking, social service classifications and clustering, social service migrations, social service composition, social service recommendations, security and privacy are main research directions in social clouds. Some open issues in this regard are as follows:

- How to use and analyze social media data to improve service quality in clouds?

- Visualized and semantic cloud customer support via social media?
- Relationship mining via social media for privacy and security problems in clouds.
- Cloud deployed event detection via social media analysis.
- Value and knowledge creation through cloud deployed social media analytical tools in wide variety of domains.

III. CONCLUSION

The basic idea of cloud computing has been proposed in 1966 by McCarthy. Salesforce.com has announced its first cloud hosted service in 1999. Public, Private, Hybrid and Community clouds are deployment models of cloud computing. Infrastructure as a Service (IaaS), Platform as a Service (PaaS) and Software as a Service (SaaS) are service models of cloud computing. There are over 65 definitions for cloud computing published by researchers and scientists. After reviewing most of definitions, authors of the paper defined cloud computing as transparent, scalable and easily accessible system with various service levels in public and/or private forms which provides on demand access to virtualized pool of resources and targets cost reductions in computing and improvements in software deployment process and IT solutions. Research challenges and directions have been categories into 6 groups namely Security, Autonomic Resource Management (ARM), Cloud Adoption, Cloud development and benchmarking big data technologies and cloud computing and social clouds. Security is the most common challenge of cloud computing. This paper is the outcome of literature review of over 190 resources including books, journal and conference papers, research project reports and deliverables, European Commission roadmaps and calls for proposals, online weblogs, white papers, business reviews, company websites, profiles and offered services by cloud providers.

REFERENCES

[1] Abelson, Hal, ed (1999). Architects of the Information Society, Thirty-Five Years of the Laboratory for Computer Science at MIT. MIT Press. ISBN 978-0262071963.
[2] David Cearley and Gene Phifer, "Case Studies in Cloud Computing", Webinar hosted by Gartner, March 2010, http://www.gartner.com/it/content/1286700/1286717/march_4_case_studies_in_cloud_computing_dcearley_gphifer.pdf.
[3] Douglas Parkhill., The Challenge of the Computer Utility, Addison-Wesley, 1966.
[4] R. Buyya, ed., High Performance Cluster Computing, vols. 1 and 2, Prentice Hall, Old Tappan, N.J., 1999.
[5] W. Dawoud, I. Takouna, and C. Meinel, "Infrastructure as a service security: Challenges and solutions," in Informatics and Systems (INFOS), 2010 The 7th International Conference on, 2010, pp. 1–8.
[6] Bohlouli, M. & Analoui, M., Grid-HPA: Predicting Resource Requirements of a Job in the Grid Computing Environment, WASET, 2008, 45, 747-751.
[7] Chen, Y.Y. Chung, C.Y. Luo, F.J. Dong, Z.Y. Wong, K.P. "An Overview of Grid Computing Application for Distributed Heterogeneous Power Systems: History, Technologies, and Prospect", 8th International Conference on Advances in Power System Control, Operation and Management (APSCOM 2009).
[8] Gartner Group. Gartner's hype cycle for emerging technologies, 2011. Gartner Group, July 2011. Visited on October 2012 via http://www.gartner.com/.
[9] BitNami, Cloud.com and Zenoss, Special Report on Cloud Computing Outlook Survey, 2011. Available via: http://cloud.com/cloud-computing-outlook
[10] North Bridge, GigaOM Pro, & 451 Group, The 2011 Future of Cloud Computing Survey, Technical Report, Available via http://www.northbridge.com/future-cloud-survey-shows-significance-open-source.
[11] The association of data center management professionals, AFCOM, 2011 State of the Data Center Report Available via http://www.pcworld.idg.com.au/article/385766/data_center_double_duty/
[12] Caraher, K., Nott, M., From Tactic to Strategy: The CDW 2011 Cloud Computing Tracking Poll, 2011.
[13] Reston, V., IT Operations in Flux as Cloud Computing Adoption Increases According to ScienceLogic Survey (2011), An online Report accessed via: http://www.sciencelogic.com/news-and-events/press-releases/it-operations-flux-cloud-computing-adoption-increases-according
[14] S. Reid and H. Kisker, "sizing the cloud- a bt futures report", Forrester 2011.
[15] Livesey, K., Public cloud services market to hit revenues of $66 billion, Ovum's Public Cloud Services Global Market Forecast Model, 2011.
[16] Mullins, R., Businesses Have Tough Decisions To Make About Cloud Computing, Network Computing, June 2011.
[17] S. Subashini and V. Kavith, "A Survey of security issues in service delivery models of cloud computing", Journal of Network and Computer Applications, Volume 34, Issue 1, January 2011, Pages 1-11.
[18] Gantz, J.F. The 2011 Digital Universe Study: Extracting Value from Chaos, IDC, sponsored by EMC, May 2011.
[19] Bohlouli, M.; Merges, F. & Fathi, M., A Cloud-based Conceptual Framework for Knowledge Integration in Distributed Enterprises, International Conference on Electro/Information Technology 2012 (IEEE EIT 2012), 2012.
[20] S. Pearson, "Taking account of privacy when designing cloud computing services," in Software Engineering Challenges of Cloud Computing, 2009. CLOUD'09. ICSE Workshop on, 2009, pp. 44–52.
[21] D. M. Rousseau, S. B. Sitkin, R. S. Burt, and C. Camerer, "Not so different after all: A cross-discipline view of trust.," Academy of management review, vol. 23, no. 3, pp. 393–404, 1998.
[22] S. Bradshaw, C. Millard, and I. Walden, "Contracts for clouds: comparison and analysis of the Terms and

Conditions of cloud computing services," International Journal of Law and Information Technology, vol. 19, no. 3, pp. 187–223, 2011.
[23] A. Sirisha and G. G. kumari, "API access control in cloud using the Role Based Access Control Model," presented at the Trends in Information Sciences & Computing (TISC), 2010, pp. 135 – 137.
[24] D. Hubbard and M. Sutton, "Top Threats to Cloud Computing," Cloud Security Alliance. Cloud Security Alliance, 2010.
[25] D. Zissis and D. Lekkas, "Addressing cloud computing security issues," future generation computer systems, vol. 28, no. 3, pp. 583-592, Mar. 2012.
[26] M. Maurer, I. Brandic and R. Sakellariou, "Simulating Autonomic SLA Enactment in Clouds using Case Based Reasoning," ServiceWave 2010, 13th-15th December 2010, Ghent, Belgium.
[27] V. C. Emeakaroha, M. A. S. Netto, R. N. Calheiros, I. Brandic, R. Buyya, and C. A. F. De Rose, "Towards autonomic detection of sla violations in cloud infrastructures," Future Generation Computer Systems, 2011.
[28] A. Gulati, G. Shanmuganathan, A. Holler, and I. Ahmad, "Cloud-Scale Resource Management: Challenges and Techniques," Architecture, 2010.
[29] R. Buyya, A. Beloglazov, and J. Abawajy, "Energy-Efficient Management of Data Center Resources for Cloud Computing: A Vision, Architectural Elements, and Open Challenges," Proceedings of the 2010 International Conference on Parallel and Distributed Processing Techniques and Applications (PDPTA 2010), Las Vegas, USA, July 12-15, 2010. - Keynote Paper.
[30] S. S. Wang, K. Q. Yan, and S. C. Wang, "Achieving efficient agreement within a dual-failure cloud-computing environment," Expert Systems with Applications, vol. 38, no. 1, pp. 906–915, 2011.
[31] W. Zhao, P. M. Melliar-Smith, and L. E. Moser, "Fault tolerance middleware for cloud computing," in Cloud Computing (CLOUD), 2010 IEEE 3rd International Conference on, 2010, pp. 67–74.
[32] PL. Bannerman, "Cloud Computing Adoption Risks: State of Play," in 17th Asia Pacific Software Engineering Conference Cloud Workshop, 2010, pp. 10–16.
[33] D. Greenwood, A. Khajeh-Hosseini, J. W. Smith, I. Sommerville: "The Cloud Adoption Toolkit: Addressing the Challenges of Cloud Adoption in Enterprise," CoRR abs/1003.3866: 2010.
[34] B. F. Cooper, A. Silberstein, E. Tam, R. Ramakrishnan, R. Sears: "Benchmarking Cloud Serving Systems with YCSB". SoCC 2010.
[35] J.Schad, J. Dittrich, J. Arnulfo Quiané-Ruiz: "Runtime Measurements in the Cloud: Observing, Analyzing,and Reducing Variance." VLDB 2010.
[36] Donald Kossmann, Tim Kraska: Data Management in the Cloud: Promises, State-of-the-art, and Open Questions. Datenbank-Spektrum 10(3): 121-129 (2010).
[37] M. A. El-Refaey and M. A. Rizkaa, "CloudGauge: a dynamic cloud and virtualization benchmarking suite," in Enabling Technologies: Infrastructures for Collaborative Enterprises (WETICE), 2010 19th IEEE International Workshop on, 2010, pp. 66–75.
[38] I. Livenson and E. Laure. Towards Transparent Integration of Heterogeneous Cloud Storage Platforms. In The Fourth International Workshop on Data Intensive Distributed Computing (DIDC 2011), San Jose, USA, June 2011.
[39] B. J. Lheureux, D. C. Plummer, T. Bova, M. Cantara, E. Knipp, P. Malinverno, Who's Who in Cloud Services Brokerage, Gartner Research, Technical Report, 2011.
[40] D. Agrawal, S. Das, A. El Abbadi, Big Data and Cloud Computing: Current State and Future Opportunities, IEEE EDBT 2011, March 22–24, 2011, Uppsala, Sweden
[41] S. K. Nair, S. Porwal, T. Dimitrakos, A. J. Ferrer, J. Tordsson, T. Sharif, C. Sheridan, M. Rajarajan, and A. U. Khan, "Towards Secure Cloud Bursting, Brokerage and Aggregation," in Web Services (ECOWS), 2010 IEEE 8th European Conference on, 2010, pp. 189–196.
[42] J. Catone, How Much Data Will Humans Create & Store This Year? [INFOGRAPHIC], Mashable Social Media, June 2011.
[43] D. Webster, IP Traffic to Quadruple by 2015, Cisco Blog Post, June 2011.
[44] Eric A. Brewer. Towards robust distributed systems. (Invited Talk) Principles of Distributed Computing, Portland, Oregon, July 2000
[45] H. Demirkan and D. Delen, "Leveraging the capabilities of service-oriented decision support systems: Putting analytics and big data in cloud," Decision Support Systems, 2012.
[46] C. N. Höfer and G. Karagiannis, "Cloud computing services: taxonomy and comparison," Journal of Internet Services and Applications, pp. 1–14, 2011.
[47] M. Bohlouli, F. Schulz, L. Angelis, D. Pahor, I. Brandic, D. Atlan, R. Tate, Towards an Integrated Platform for Big Data Analysis, International Conference on Integrated Systems Design and Technology (ISDT12), 2012, Spain.
[48] S. Chaudhuri, "What next?: a half-dozen data management research goals for big data and the cloud," in Proceedings of the 31st symposium on Principles of Database Systems, 2012, pp. 1–4.
[49] List of NoSQL databases, visited on October 2012 via http://nosql-database.org/
[50] R. Cattell, "Scalable sql and nosql data stores," ACM SIGMOD Record, vol. 39, no. 4, pp. 12–27, 2011
[51] Large Scale Dtat Analysis on Cloud Systems, Available via http://ercim-news.ercim.eu/en89/special/large-scale-data-analysis-on-cloud-systems
[52] S. Li, and Z. Chen, "Social service computing: concepts, research, challenges and directions," 2010 IEEE/ACM international conference on green computing and communications & 2010 IEEE/ACM international conference on cyber,physical and social computing, 2010, pp. 840-845
[53] P. Russom, "Big Data Analytics," TDWI Best Practices Report, 4 th Quarter 2011, 2011.
K. Chard, S. Caton, O. Rana, and K. Bubendorfer, "Social cloud: Cloud computing in social networks," in Cloud Computing (CLOUD), 2010 IEEE 3rd International Conference on, 2010, pp. 99–106.